\documentclass[12pt,preprint]{aastex}
\usepackage{color}

\begin{document}

\title{Beaming and rapid variability of high-energy radiation from relativistic pair plasma reconnection}

\shorttitle{Beaming and rapid variability of high-energy radiation from relativistic pair plasma reconnection}

\author{B.~Cerutti$^{1}$, G.~R.~Werner$^{1}$, D.~A.~Uzdensky$^{1}$ \& M.~C.~Begelman$^{2,3}$} \shortauthors{Cerutti, Werner, Uzdensky, \& Begelman}

\affil{$^1$ Center for Integrated Plasma Studies, Physics Department, University of Colorado, UCB 390, Boulder, CO 80309-0390, USA; benoit.cerutti@colorado.edu, greg.werner@colorado.edu, uzdensky@colorado.edu}

\affil{$^2$ JILA, University of Colorado and National Institute of Standards and Technology, UCB 440, Boulder, CO 80309-0440, USA; mitch@jila.colorado.edu}

\affil{$^3$ Department of Astrophysical and Planetary Sciences, University of Colorado, UCB 391, Boulder, CO 80309-0391, USA}

\begin{abstract}
We report on the first study of the angular distribution of energetic particles and radiation generated in relativistic collisionless electron-positron pair plasma reconnection, using two-dimensional particle-in-cell simulations. We discover a strong anisotropy of the particles accelerated by reconnection and the associated strong beaming of their radiation. The focusing of particles and radiation increases with their energy; in this sense, this ``kinetic beaming'' effect differs fundamentally from the relativistic Doppler beaming usually invoked in high-energy astrophysics, in which all photons are focused and boosted achromatically. We also present, for the first time, the modeling of the synchrotron emission as seen by an external observer during the reconnection process. The expected lightcurves comprise several bright symmetric sub-flares emitted by the energetic beam of particles sweeping across the line of sight intermittently, and exhibit super-fast time variability as short as about one tenth of the system light-crossing time. The concentration of the energetic particles into compact regions inside magnetic islands and particle anisotropy explain the rapid variability. This radiative signature of reconnection can account for the brightness and variability of the gamma-ray flares in the Crab Nebula and in blazars.
\end{abstract}


\section{Introduction}\label{intro}

The high-energy radiation from numerous astrophysical objects, including active galactic nuclei, pulsar wind nebulae, and gamma-ray bursts, is emitted by particles accelerated to relativistic speeds. Magnetic reconnection is one of the main mechanisms thought to accelerate particles, by converting magnetic energy into particle kinetic energy (see, e.g., the review by \citealt{2009ARA&A..47..291Z}). Previous numerical Particle-In-Cell (PIC) studies of reconnection in relativistic electron-positron pair plasmas (e.g., \citealt{2001ApJ...562L..63Z, 2007ApJ...670..702Z, 2008ApJ...677..530Z, 2004PhPl...11.1151J, 2007PhPl...14e6503B, 2012ApJ...750..129B, 2007A&A...473..683P, 2009PhRvL.103g5002J, 2011PhPl...18e2105L, 2011ApJ...741...39S}) provide a detailed picture of the particle energy spectrum. However, by itself, the energy spectrum lacks information critical to the determination of radiation generated by the plasma --- namely, the angular distribution of the velocities of accelerated particles. Because ultra-relativistic particles radiate in a narrow cone along their direction of motion, any anisotropy of the energetic particles translates directly into the anisotropy of their emission (e.g., synchrotron or inverse Compton). The beaming of the radiation drastically affects how we infer, from observations, the physical conditions of the emitting region (e.g., size, overall energetics and dynamics) and statistical properties of flaring astrophysical objects.

In this Letter, we report on the first detailed analysis of the angular distribution (in addition to the energy and spatial distributions) of particles accelerated in collisionless relativistic pair reconnection, using PIC simulations. In Section~\ref{setup} we describe the simulation setup. In Section~\ref{res}, we present our results and report on the discovery of a ``kinetic beaming'', i.e., a strong energy-dependent anisotropy of the particles and their radiation. We also present, for the first time, the modeling of the high-energy radiation spectrum and lightcurve as seen by a distant observer, and predict extremely rapid time-variability (much shorter than the light-crossing time of the system). In Section~\ref{ccl}, we briefly discuss the general implications of our findings in the astrophysical context of flaring high-energy gamma-ray sources like the Crab Nebula and blazars.

\section{PIC simulation setup}\label{setup}

We performed two-dimensional numerical simulations of collisionless relativistic pair plasma reconnection using the explicit electromagnetic PIC capabilities of {\tt vorpal} \citep{2004JCoPh.196..448N}. The initial setup adopted here is standard in reconnection simulations (see, e.g., \citealt{2001ApJ...562L..63Z}). It consists of a rectangular box of size $L_{\rm x}\times L_{\rm y}$ with two anti-parallel relativistic Harris current layers \citep{2003ApJ...591..366K} and double periodic boundary conditions. In the following, we will focus on the dynamics of the bottom layer only, i.e., the bottom half of the simulation domain. The reconnecting magnetic field is $B_{\rm x}=B_0\tanh(y/\delta)$, where $B_0$ is the upstream field and $\delta$ is the initial layer thickness. There is no guide field, $B_{\rm z}=0$. The simulation has a resolution of $4.6$ grid cells per $\delta\approx 0.8\rho_{\rm c}$, where $\rho_{\rm c}=m_{\rm e}c^2/eB_0$ is the non-relativistic electron Larmor radius, $m_{\rm e}$ is the electron rest mass, $e$ the elementary electric charge, and $c$ the speed of light.

The initial particle density (electrons and positrons together in the laboratory frame) is $n=n_{\rm drift}+n_0$, where $n_{\rm drift}=n_{\rm d0}\cosh^{-2}(y/\delta)$ is the density of electrons and positrons drifting in opposite directions (in the $\pm z$-direction) at a velocity $v_{\rm drift}/c=0.6$ and located in the layer, and $n_0=0.042 n_{\rm d0}$ is a uniform and isotropic background density. Both populations are distributed according to a relativistic Maxwellian of temperature $kT'_{\rm drift}=0.3 m_{\rm e}c^2$ defined in the drifting particles co-moving frame, and $kT_{\rm bg}=0.15 m_{\rm e}c^2$ in the laboratory frame, where $k$ is the Boltzmann constant. The electron skin depth in the layer is $d_{\rm e}=c/\omega_{\rm pe}\approx 0.7\rho_{\rm c}$, where $\omega_{\rm pe}=\sqrt{4\pi n_{\rm d0}e^2/m_{\rm e}}$ is the plasma frequency defined with the drifting electrons and positrons. The upstream plasma beta parameter is $\beta=8\pi n_{\rm 0} k T_{\rm bg}/B_0^2\approx 2.6\times 10^{-2}$ (the magnetization sigma parameter is $\sigma=B_0^2/4\pi n_0 m_{\rm e}c^2\approx 11.5$). Radiative energy losses are neglected (see \citealt{2009PhRvL.103g5002J} for a PIC simulation including radiative drag).

\begin{figure}
\epsscale{0.75}
\plotone{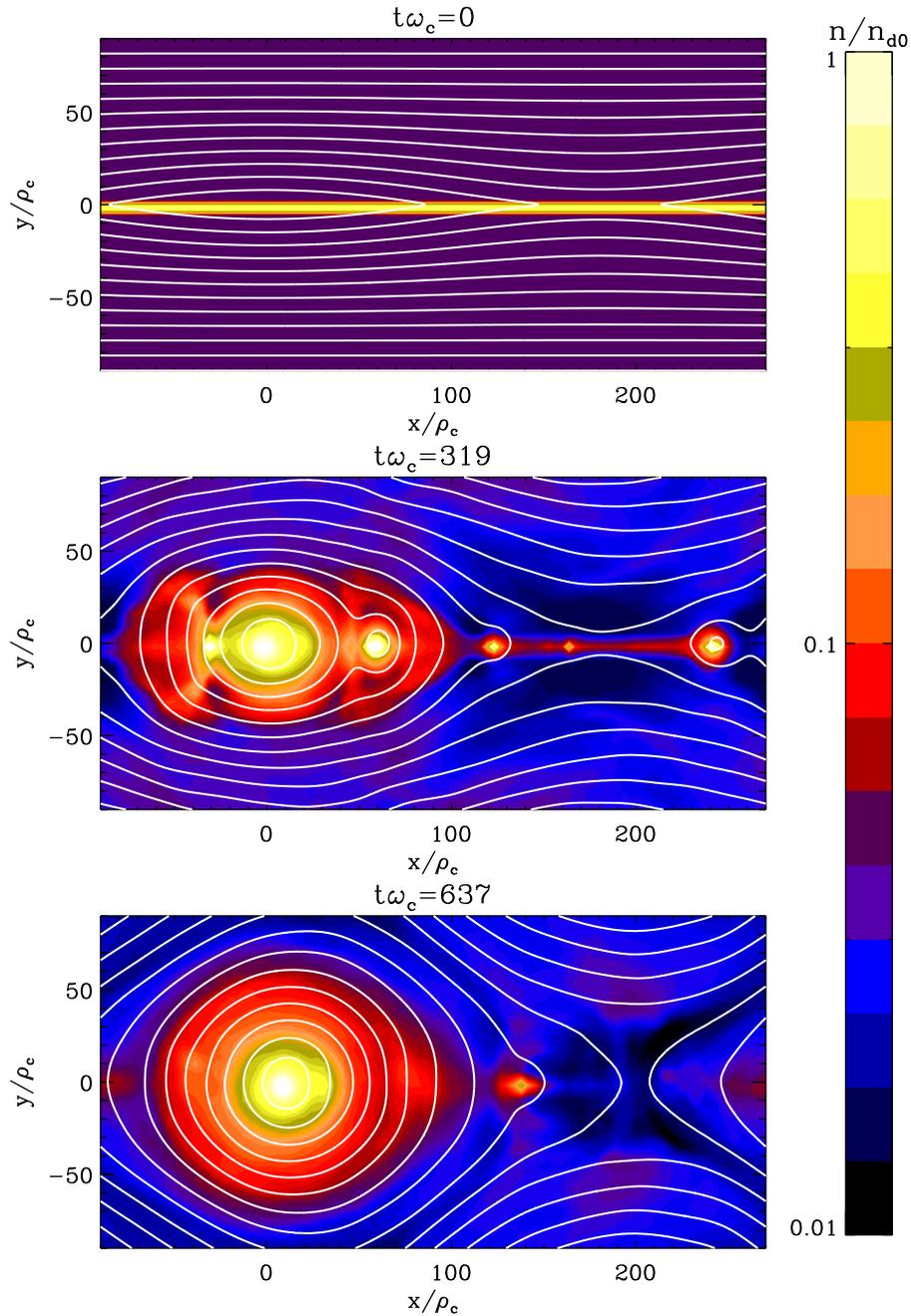}
\caption{Plasma density map and magnetic field lines (white solid lines) at the initial stage $t\omega_{\rm c}=0$ (top panel), at the intermediate stage $t\omega_{\rm c}=319$ (middle panel), and close to the final stage of the reconnection event $t\omega_{\rm c}=637$ (bottom panel). Only the bottom half of the simulation box is shown here. Distances in the $x$- and $y$-directions are normalized to the non-relativistic electron Larmor radius $\rho_{\rm c}$. The density is normalized to the initial drifting particle density $n_{\rm d0}$.}
\label{cerutti_fig1}
\end{figure}

The unit of length in the $x$- and $y$-directions is $\rho_{\rm c}$, and the unit of time is the inverse of the nominal cyclotron frequency $\omega^{-1}_{\rm c}=\rho_{\rm c}/c$. In order to initiate the reconnection process, we break the initial Harris equilibrium with a tearing-like perturbation in the magnetic flux function (Figure~\ref{cerutti_fig1}, top panel). We performed a series of simulations with different box sizes $(L_{\rm x}/\rho_{\rm c},L_{\rm y}/\rho_{\rm c})=(90,90),~(180,180),~(360,360),$ and $(720,720)$, and with $16$, $64$, and $256$ particles per cell. We verified numerical convergence with respect to both the spatial resolution (i.e., the number of grid cells per $\rho_{\rm c}$) and the number of particles per cell. The results shown below are for a $L_{\rm x}=L_{\rm y}=360\rho_{\rm c}$ and $2.7\times 10^8$ particles ($64$ per cell).

\section{Results}\label{res}

\subsection{Overall reconnection dynamics}\label{dyn}

Figure~\ref{cerutti_fig1} shows the three main stages of the reconnection dynamics for the bottom layer only. The initial current layer quickly becomes unstable to secondary tearing modes (e.g., \citealt{2004PhPl...11.1151J,2007PhPl...14j0703L,2007MNRAS.374..415K,2011PhPl...18e2105L}), breaking into a dynamical chain of magnetic islands made of closed magnetic flux loops, visible in the intermediate stage at $t=319\omega^{-1}_{\rm c}\approx 0.9 L_{\rm x}/c$ (middle panel in Figure~\ref{cerutti_fig1}), when about 44\% of the initial magnetic energy has been dissipated. The number of islands subsequently decreases as small islands merge into bigger ones. We estimate the time-averaged dimensionless reconnection rate to be $\beta_{\rm rec}=E_{\rm z}/B_0\approx 0.16$ (where $E_{\rm z}$ is the electric field in the $z$-direction), up to $t=600\omega^{-1}_{\rm c}$ when most of the field has reconnected. At the end of the reconnection event, the system settles in a saturated configuration with a single big island, or O-point, and a single X-point (see bottom panel in Figure~\ref{cerutti_fig1}, $t=637\omega^{-1}_{\rm c}$) in each half of the box. The final state in the upper-half of the simulation domain (not shown here) is identical to the bottom-half but symmetric with respect to the center of the box. All the magnetic flux ends up around the two main islands, and the separatrices emanating from the two X-points connect to each other. During this process, about 55\% of the initial magnetic energy is converted into particle kinetic energy. The total energy in the system is conserved with less than 0.1\% of error at the end of the simulation $t=1270\omega^{-1}_{\rm c}$.

\begin{figure}
\epsscale{0.65}
\plotone{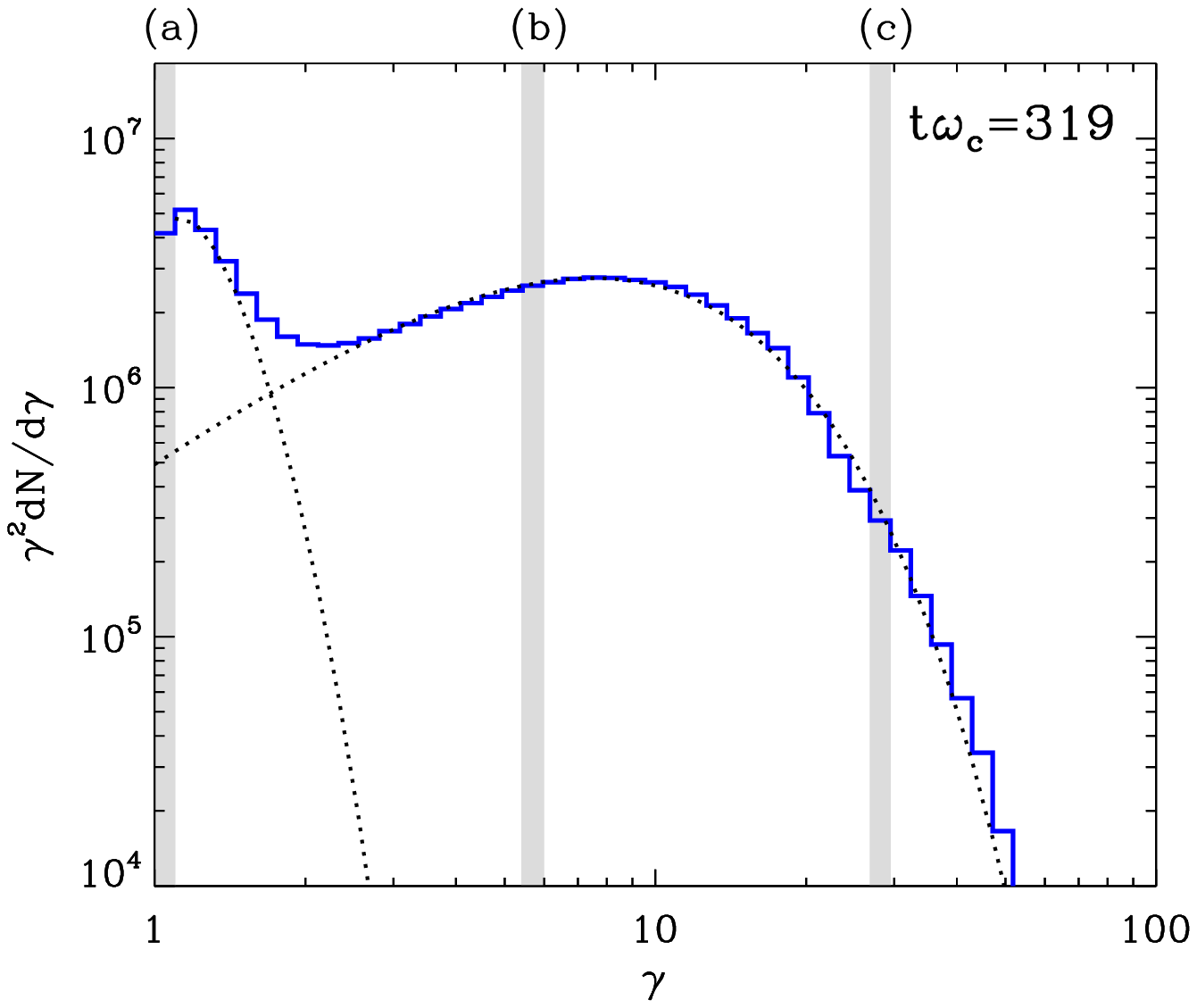}
\plotone{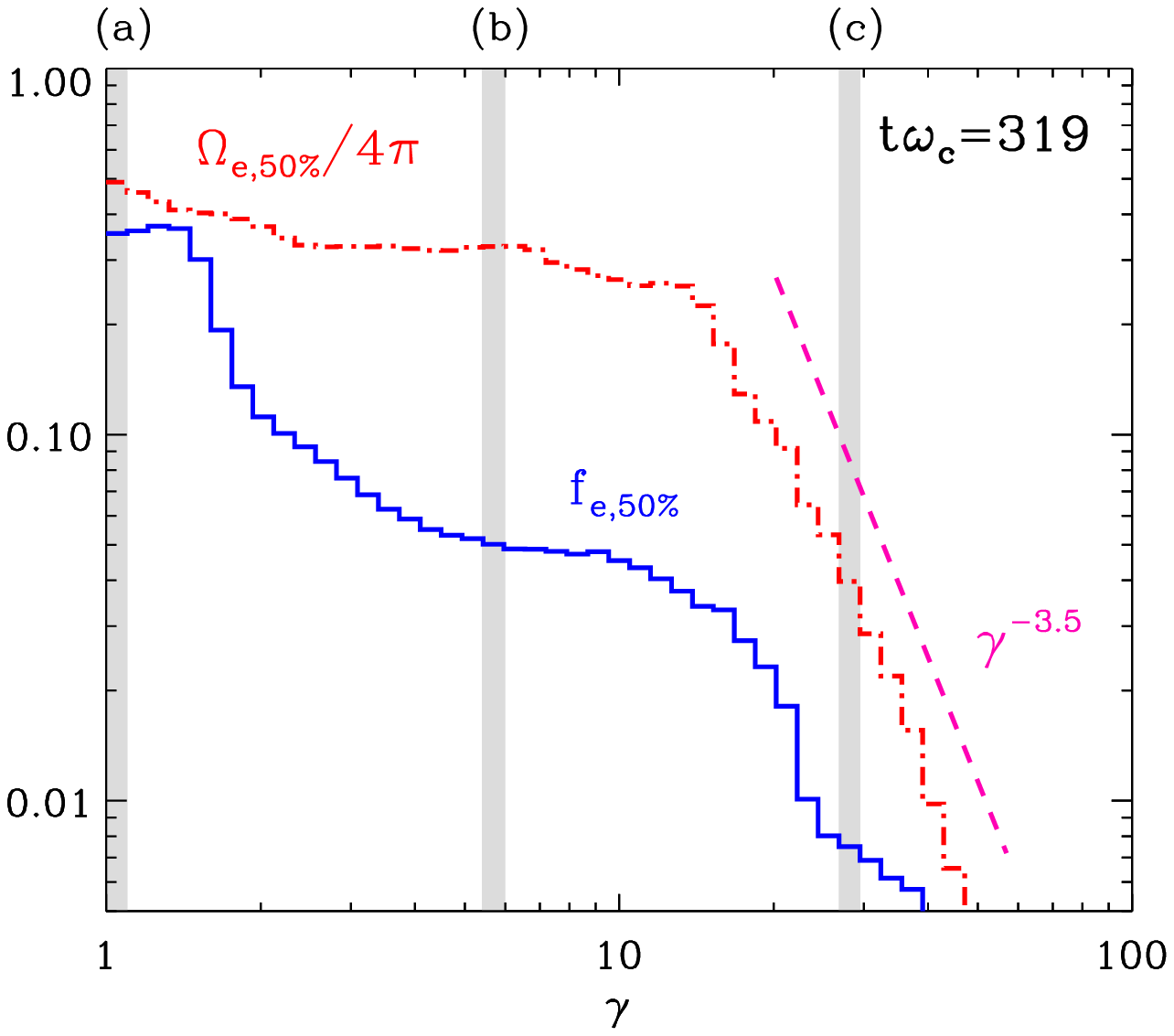}
\caption{Top panel: Energy distribution function $\gamma^2 dN/d\gamma$ of all the positrons in the simulation box (blue solid line) as function of $\gamma$ at $t\omega_{\rm c}=319$. The dotted lines are analytical fits to the low- (relativistic Maxwellian of temperature $kT=0.15 m_{\rm e}c^2$) and high-energy ($dN/d\gamma\propto\gamma^{-1/2}\exp(-\gamma/5)$) parts of the energy distribution function. Bottom panel: Solid angle normalized by $4\pi$, $\Omega_{{\rm e},50\%}/4\pi$ (red dot-dashed line), and spatial filling factor, $f_{{\rm e},50\%}$ (blue solid line), containing half of the positrons in a given energy bin, as functions of $\gamma$ at $t\omega_{\rm c}=319$. The pink dashed line is a power-law fit to $\Omega_{{\rm e},50\%}/4\pi$ for $\gamma>20$. The gray bands are three particle energy bins (a), (b), and (c) for which the angular and spatial distributions of the particles are shown in Figure~\ref{cerutti_fig3}.}
\label{cerutti_fig2}
\end{figure}

\begin{figure*}
\epsscale{1.0}
\plotone{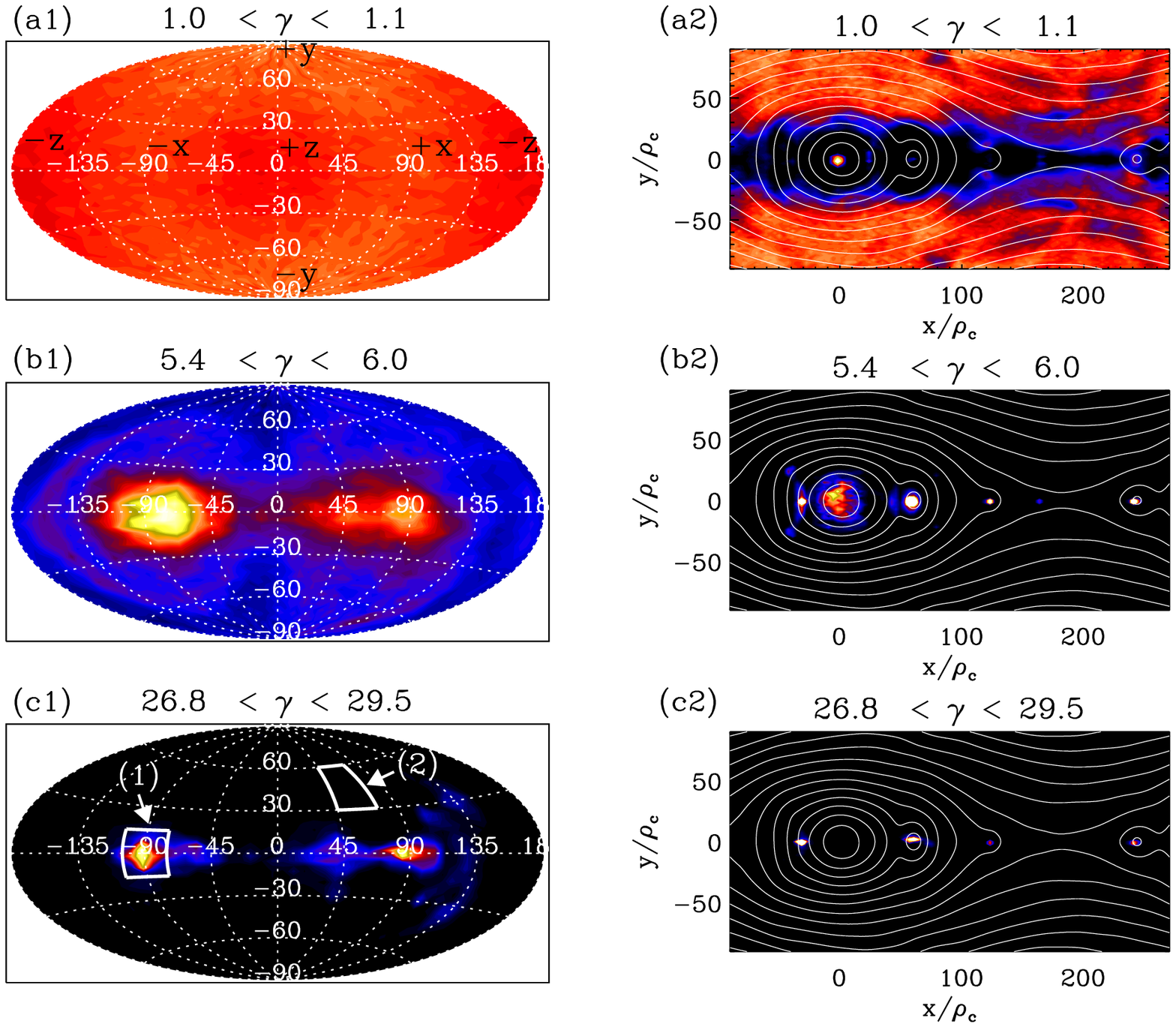}
\caption{Angular distribution maps (using the Aitoff projection, panels (a1)-(c1)), and spatial distribution maps (panels (a2)-(c2)) of the positrons in the three energy bands (a)-(c) defined in Figure~\ref{cerutti_fig2}. In the angular maps, the $x$-axis gives the value for the longitude $\lambda$ and the $y$-axis the latitude $\phi$ (see the text for their definitions). In panel (a1), the directions $\pm x$, $\pm y$ and $\pm z$ are indicated. The subdomains $-15\degr<\phi<+15\degr$, $-105\degr<\lambda<-75\degr$ labeled ``(1)'' and $+30\degr<\phi<+60\degr$, $+45\degr<\lambda<+75\degr$ labeled ``(2)'' in panel (c1) shown by the white arrows and delimited by white squares are used in Figure~\ref{cerutti_fig4} to compute the synchrotron radiation spectrum emitted in these specific directions. Bright/dark colors show high/low densities of particles per unit of solid angle (left panels) and per unit of area (right panels).}
\label{cerutti_fig3}
\end{figure*}

\subsection{Particle energy distribution, particle anisotropy and spatial inhomogeneity}\label{anis}

Shortly after the onset of reconnection, a new population of relativistic particles emerges from the initial cool thermal distribution. Particles are accelerated along the $z$-direction at X-points by the strong reconnection electric field $E_{\rm z}$. They are then deflected along the $\pm x$-direction by the reconnected magnetic field $B_{\rm y}$ \citep{2001ApJ...562L..63Z,2011ApJ...741...39S}. Figure~\ref{cerutti_fig2} (upper panel) presents the energy distribution $\gamma^2 dN/d\gamma$ of all the positrons in the simulation box (blue solid line), where $\gamma$ is the particle Lorentz factor, at the intermediate stage of reconnection $t=319\omega^{-1}_{\rm c}$. The high-energy bump peaking at $\gamma\approx 8$ can be well fitted by $dN/d\gamma\propto\gamma^{-1/2}\exp(-\gamma/5)$. We describe this component as a quasi-thermal distribution resulting from plasma heating by magnetic dissipation, rather than a non-thermal power-law tail. It contains about $49\%$ of the particles and $78\%$ of the kinetic energy.

The main result of this paper concerns the energetic particles' anisotropy, which we examine by calculating the total solid angle within which half of the particles of a given energy are contained, $\Omega_{\rm e,50\%}(\gamma)$, as a function of the particle energy (Figure~\ref{cerutti_fig2}, bottom panel). The low-energy particles ($\gamma\lesssim 10$) remain approximately isotropic, whereas the high-energy particles ($\gamma\gtrsim 20$) are focused in a tight beam whose solid angle decreases rapidly with energy, roughly as $\Omega_{\rm e,50\%}\propto \gamma^{-3.5}$. The beam's angular size can become as small as $\Omega_{\rm e,50\%}/4\pi<1\%$ of the whole sphere for $\gamma\gtrsim 40$.

In addition to the anisotropy, we study the spatial distribution of the energetic particles in the computational domain. We quantify the degree of inhomogeneity of the particles by computing the fraction of the system surface ($L_{\rm x}\times L_{\rm y}$) covered by half of the particles, as a function their energy, $f_{\rm e,50\%}(\gamma)$ (Figure~\ref{cerutti_fig2}, bottom panel). We find a strong energy-dependent inhomogeneity: the particles with $\gamma<2$ fill a large fraction of the box ($f_{\rm e, 50\%}\approx 0.4$), mostly outside magnetic islands, while the higher energy particles ($2<\gamma<20$) are concentrated into small bunches ($f_{\rm e,50\%}\lesssim 0.1$) inside islands. These results are consistent with \citet{2011PhPl...18e2105L} simulations. For $\gamma>20$, $f_{\rm e,50\%}$ drops abruptly.

Figure~\ref{cerutti_fig3} provides a good illustration of the anisotropy and inhomogeneity of the particle distribution in the three different energy bins shown in Figure~\ref{cerutti_fig2}. The angular distribution is calculated using spherical coordinates in which a radial unit vector has the coordinates $x=\cos\phi\sin\lambda$, $y=\sin\phi$, $z=\cos\phi\cos\lambda$, where $\lambda$ is the longitude and $\phi$ the latitude. After the end of reconnection, the energy dependence of anisotropy and inhomogeneity decreases because all particles end up circling inside the O-point (Figure~\ref{cerutti_fig1}, bottom panel).

\begin{figure}
\epsscale{1.0}
\plotone{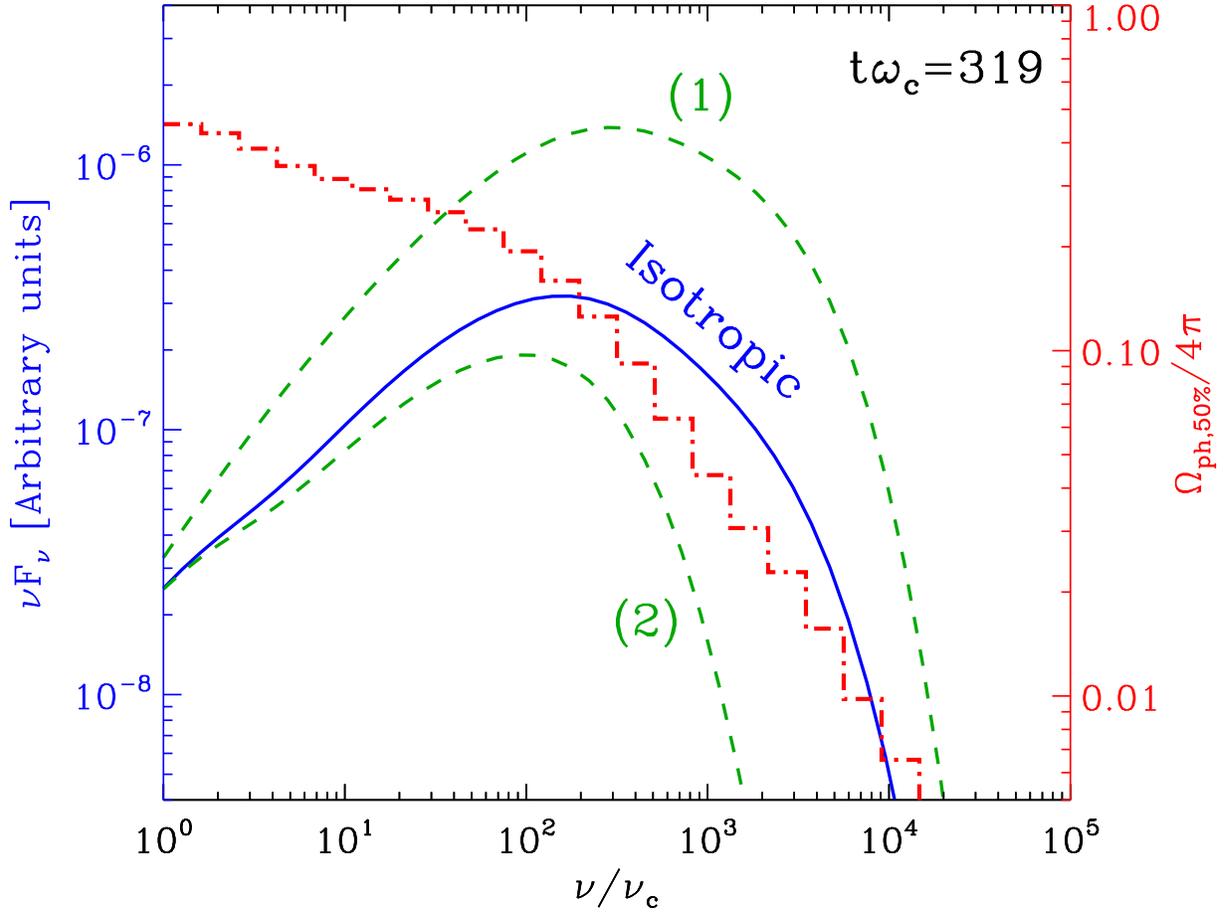}
\caption{Instantaneous spectral energy distribution emitted by all the positrons in the bottom-half of the simulation box via synchrotron radiation $\nu F_{\nu}$ averaged over all directions as a function of the reduced frequency $\nu/\nu_{\rm c}$ (blue solid line), where $\nu_{\rm c}=3\omega_{\rm c}/4\pi$, at $t\omega_{\rm c}=319$. For comparison, the green dashed lines show the spectral energy distributions emitted by the particles contained in the solid angle domains (1) and (2) defined in Figure~\ref{cerutti_fig3}, panel (d). This figure shows also the solid angle containing half of the photons in a given frequency bin, $\Omega_{{\rm ph},50\%}$, normalized by $4\pi$, as a function of $\nu/\nu_{\rm c}$ (red dot-dashed line).}
\label{cerutti_fig4}
\end{figure}

\subsection{Synchrotron beam}\label{rad}

Next, we examine the main radiative signatures of the reconnection process, namely, the emission spectrum and temporal variability. We first look at the effect of particle anisotropy on the observable synchrotron radiation spectrum emitted by the layer. Following the same procedure as for the particle energy and angular distributions, we characterize the photon distribution emitted by the particles. Our analysis makes four approximations: (1) the particles emit pure synchrotron radiation, (2) the plasma does not absorb the radiation (optically thin), (3) all the emission is beamed in the direction of motion of the radiating particle (valid for $\gamma\gg 1$), and (4) synchrotron energy losses and the radiation reaction force on the particles are ignored. All the results presented below regarding the calculation of radiation are performed after the simulation is completed, in accordance with assumptions (2) and (4).

Using the classical synchrotron spectrum formula (e.g., \citealt{1970RvMP...42..237B}), we calculate the resulting instantaneous photon spectral energy distribution (SED, i.e., radiative power per unit of area) emitted by all the positrons in the box at $t\omega_{\rm c}=319$ (Figure~\ref{cerutti_fig4}). Frequencies are normalized to the nominal critical synchrotron frequency $\nu_{\rm c}=3\omega_{\rm c}/4\pi$. The overall shape of the SED averaged over all directions $\langle\nu F_{\nu}\rangle_{\rm iso}$ (blue solid line) resembles the shape of the particle energy distribution in Figure~\ref{cerutti_fig2}. The spectral peak coincides with the typical synchrotron photon frequency of the bulk of energetic particles ($\gamma\sim 10$), i.e. $\nu/\nu_{\rm c}\sim\gamma^2=100$. Below the peak ($\nu/\nu_{\rm c}<100$), the spectrum can be well fitted by a single power law of index $\sim +0.6$. The cool initial distribution of particles (with $kT=0.15 m_{\rm e} c^2$) is responsible for the slight flux excess at low frequencies ($\nu/\nu_{\rm c}<10$). The most energetic particles ($\gamma>10$) radiate above $\nu/\nu_{\rm c}=100$ and form a soft power-law-like component of index $\sim -0.7$ between $200<\nu/\nu_{\rm c}<2000$, followed by a sharp cut-off (see also \citealt{2004ApJ...605L...9J} for a similar calculation).

The anisotropy of the high-energy particles translates directly into the anisotropy of radiation. We compute the angular distribution of the emission using the same measures as for the particles (see Section~\ref{anis}), namely the solid angle within which half of the photons are contained, $\Omega_{\rm ph,50\%}$, as a function of $\nu/\nu_{\rm c}$ (Figure~\ref{cerutti_fig4}, red dot-dashed line). As expected, we find that the emitted flux displays a strong frequency-dependent anisotropy, very much like the particles, although the transition from the isotropic to the highly anisotropic regime, roughly at $\nu/\nu_{\rm c}=100$, is more gradual for photons. The solid angle of the radiation beam decreases with frequency approximately as $\Omega_{\rm ph,50\%}\propto(\nu/\nu_{\rm c})^{-0.75}$. The high-energy photons ($\nu/\nu_{\rm c}>100$) are concentrated in a small solid angle $\Omega_{\rm ph,50\%}/4\pi<0.1$\footnote{Although this is not the case here, if the particle beam solid angle $\Omega_{{\rm e},50\%}$ were smaller than $\sim 1/\gamma^2$, then the angular size of the radiation beam would be controlled by the opening angle of the synchrotron beam of a single particle, i.e., $\Omega_{\rm ph,50\%}\sim 1/\gamma^2$.}. The angular distribution maps (similar to Figure~\ref{cerutti_fig3}, not shown here) indicate that the high-energy radiation is strongly beamed towards the $\pm x$-directions at $t\omega_{\rm c}=319$, although the beam is changing direction restlessly within the plane of the layer during reconnection.

To illustrate the significance of beaming, we present the spectrum of photons $\langle\nu F_{\nu}\rangle_{(1)}$ emitted in the direction of the most energetic particles, e.g., around the $-x$-direction ($-15\degr<\phi<+15\degr$, $-105\degr<\lambda<-75\degr$ corresponding to $\Delta\Omega_{(1)}=0.27~$sr, see Figure~\ref{cerutti_fig3}, domain ``$(1)$'' in panel (d)), and compare it with $\langle\nu F_{\nu}\rangle_{\rm iso}$ (Figure~\ref{cerutti_fig4}). The spectrum $\langle\nu F_{\nu}\rangle_{(1)}$ is notably harder than $\langle\nu F_{\nu}\rangle_{\rm iso}$ at all frequencies. The beaming of the most energetic particles concentrates their synchrotron radiation into a small solid angle, yielding a flux more than an order of magnitude greater than the isotropic flux at the same frequency. In contrast, the observed high-energy emission is strongly suppressed in other directions as, for instance, in the solid angle domain (2) (Figures~\ref{cerutti_fig3}-\ref{cerutti_fig4}). The results are qualitatively identical for particles radiating predominantly via inverse Compton scattering, because target photons are scattered and focused in the direction of motion of the particles.

\begin{figure}
\epsscale{1.0}
\plotone{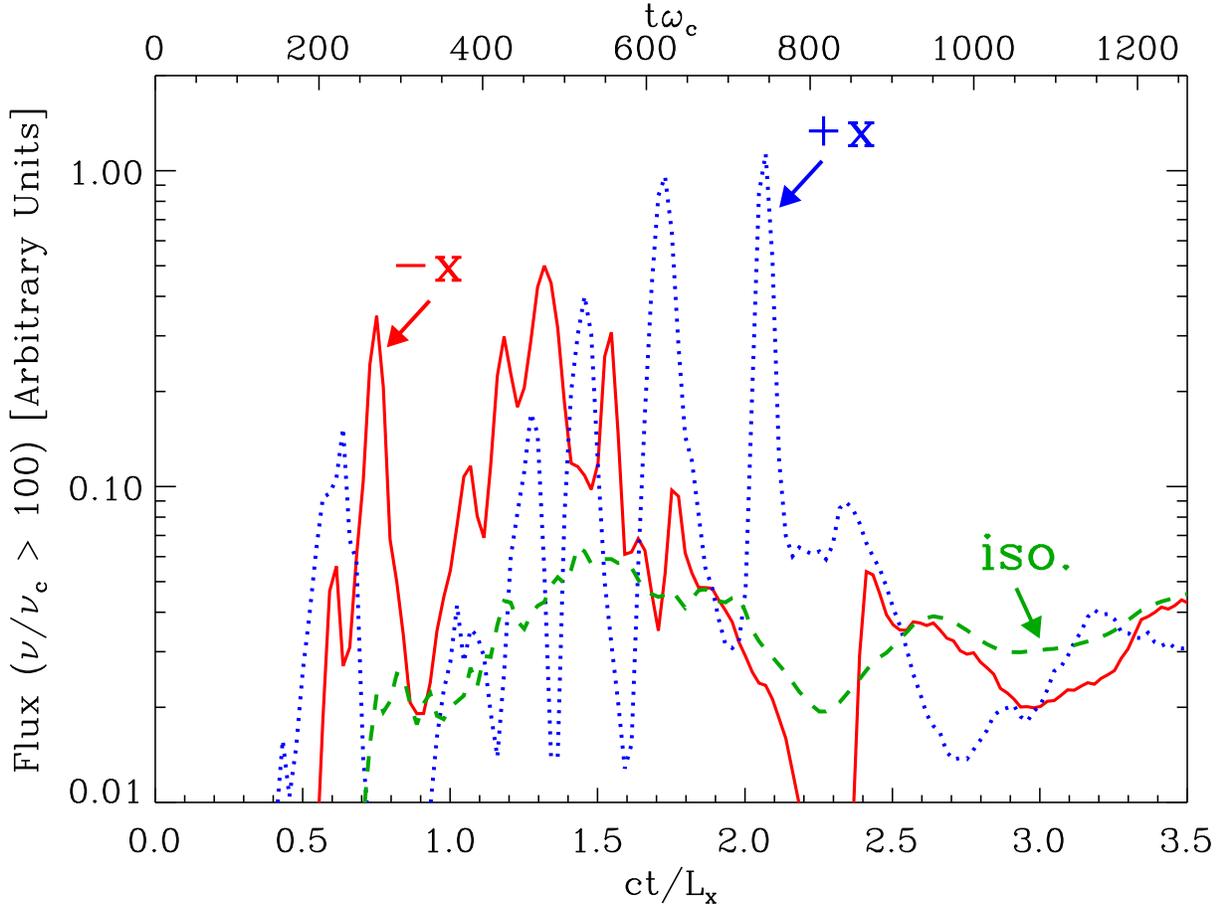}
\caption{High-energy synchrotron flux above $\nu/\nu_{\rm c}=100$ as a function of time (in units of the light-crossing time of the simulation box $L_{\rm x}/c$) seen by an observer located at infinity, looking in the $-x$ (red solid line) and $+x$-directions (blue dotted line) within $\Delta\Omega=0.03$~sr. The total lightcurve averaged over all directions is shown for comparison (green dashed line labeled ``iso.'').}
\label{cerutti_fig5}
\end{figure}

\subsection{High-energy radiation lightcurve}

Consider an observer at infinity looking in the same direction during the entire reconnection event. What would be the high-energy radiation flux seen by the observer as a function of time? To calculate the lightcurve, we compute the flux received by the observer taking into account the time delay due to the finite time of propagation of the radiation through the box. As an example, we consider an observer looking in the directions $\pm x$ ($\lambda=\pm 90\degr$, $\phi=0\degr$). We sum the contributions from all the particles going in the direction delimited by the finite but small solid angle domain $\Delta\Omega_{\pm x}=0.03~$sr centered around the $\pm x$-directions. Figure~\ref{cerutti_fig5} gives the observed photon flux integrated above $\nu/\nu_{\rm c}=100$ as a function of time.

We find that reconnection generates bright sub-flares on timescales of order one tenth the light-crossing time of the system ($L_{\rm x}/c$). The amplitude of the spikes increases with the observed radiation frequency. The short time-variability is due to the bunching of the high-energy particles into small volumes inside the magnetic islands moving away from the X-points along the $\pm x$-directions, and particle anisotropy. The high-energy beam of particles sweeps across the line of sight intermittently and generates each bright spike of the lightcurve with nearly symmetric profile (i.e., the rising time is of order the decaying time). The intense sub-flares are smoothed out if one considers the total flux averaged over all directions (Figure~\ref{cerutti_fig5}, green dashed line), demonstrating that they are caused by a geometric effect (sweeping beam) rather than an intrinsic change in the acceleration mechanism. At the end of the reconnection process, even the high-energy variability decays due to the isotropization of particles at the O-point.

\section{Astrophysical implications}\label{ccl}

The anisotropy of the particle distribution function discovered in this study leads to a strong beaming of the radiation emitted during a reconnection event. This ``kinetic beaming'' is energy-dependent, i.e., the collimation of particles and radiation increases with their energy. Kinetic beaming differs from the relativistic Doppler beaming usually invoked in high-energy astrophysics \citep{1966Natur.211..468R}: Doppler beaming is caused by the bulk motion of a plasma emitting isotropically in its rest frame; in contrast to kinetic beaming, Doppler beaming focuses and boosts all photons by the same factor regardless of their energies. This fundamental difference provides a way to discriminate observationally between these two beaming mechanisms. In addition, we expect rapid variability of the observed flux much shorter than the light-crossing time of the system with nearly symmetric burst profiles (particle bunching and sweeping beam). This situation is often encountered in high-energy astrophysics, in objects such as, e.g., active galactic nucleus jets, or gamma-ray bursts. 

The discovery of gamma-ray flares in the Crab Nebula \citep{2011Sci...331..736T,2011Sci...331..739A} is a good example, because the shortest detected variability timescale of a few hours \citep{2011A&A...527L...4B,2012ApJ...749...26B} may be much shorter than the light-crossing time of the flaring region (days to weeks). The nearly symmetric shape of the observed sub-flares suggests that the rapid variability is due to a geometric effect. This is consistent with our findings, supporting the magnetic reconnection scenario for the origin of the flares in the nebula \citep{2011ApJ...737L..40U,2012ApJ...746..148C}, in which PeV particles are accelerated and focused in a thin fan beam in the layer. In addition, the super-fast variability (variability timescales much shorter than the light-crossing time of the supermassive black hole) observed at TeV energy gamma rays in a few blazars (PKS~$2155-304$ \citep{2007ApJ...664L..71A}, Mrk~$501$ \citep{2007ApJ...669..862A} or more recently in PKS~$1222+216$ \citep{2011ApJ...730L...8A}) is difficult to explain unless one invokes extreme jet bulk Lorentz factors $\Gamma\gtrsim 50$ (see, e.g., \citealt{2006ApJ...640..185H,2008MNRAS.384L..19B}). High-energy particle anisotropy and inhomogeneity generated by magnetic reconnection in the comoving frame can alleviate the severe constraints on the energetics and collimation of the relativistic jet inferred from TeV observations \citep{2012arXiv1202.2123N}. Finally, the beaming of the high-energy radiation is also important for the interpretation of flare statistics. Gamma-ray flares could occur repeatedly but we detect only those with emission pointing toward us.

\acknowledgements We are grateful to L.~Sironi, A.~Spitkovsky, K.~Nalewajko and the referee for valuable comments on this study. This research was supported by an allocation of advanced computing resources provided by the National Science Foundation, by NSF grant PHY-0903851, NSF grant AST-0907872 and NASA Astrophysics Theory Program grant NNX09AG02G. Numerical simulations were performed with the \textsc{vorpal} framework on the local CIPS computer cluster Verus and on Kraken at the National Institute for Computational Sciences (www.nics.tennessee.edu/). We gratefully acknowledge the contributors to \textsc{vorpal}: \url{www.txcorp.com/products/VORPAL/user\_documentation/5.2\_docs/release\_install/README.html}.

\end{document}